\documentclass[letterpaper, conference]{ieeetran}

\usepackage{balance}
\usepackage{cite}
\usepackage{float}
\usepackage{multicol,multirow}
\usepackage{makecell,booktabs}
\usepackage{hyperref}
\hypersetup{    
    colorlinks=true,
    linkcolor=black,
    anchorcolor=black,
    citecolor=black,
    filecolor=black,
    menucolor=black,
    runcolor=black,
    urlcolor=black,
}

\usepackage{subcaption} % for subplot captions
\usepackage[utf8]{inputenc}

\usepackage{graphicx}
\usepackage{amsfonts}
\usepackage{amssymb}
\usepackage{amsmath,mathtools}
\usepackage{array}
\newcolumntype{P}[1]{>{\centering\arraybackslash}p{#1}}
\newcolumntype{L}[1]{>{\raggedright\arraybackslash}p{#1}}
\newcolumntype{R}[1]{>{\raggedleft\arraybackslash}p{#1}}
\usepackage{color}

\usepackage[labelsep=period]{caption}
\usepackage{bbm}
\usepackage{bm}
\usepackage{comment}
\usepackage{forest}
\usepackage{dblfloatfix}

\usepackage{etoolbox}
\usepackage{kotex}
\usepackage{breqn}
\usepackage{svg}
\usepackage{subcaption}
\usepackage{multirow}

\newcommand{\js}[1]{\textcolor{black}{#1}}
\newcommand{\jjs}[1]{\textcolor{black}{#1}}

\title{\bf \Large
    {
    Community Energy Management System for Fast Frequency Response: A Hierarchical Control Approach
    }\vspace{-4mm}
}

\author{\IEEEauthorblockN{Joonsung Jung, Hyunjoong Kim, Hyunghwan Shin and 
Jip Kim}
\IEEEauthorblockA{Dept. of Energy Engineering,
Korea Institute of Energy Technology\\
\{jjoonstar, hyunjoongkim, gudghks0217, jipkim\}@kentech.ac.kr}
\thanks{This work was supported by National Research Foundation of Korea (NRF) grant funded by the Korean government (MIST) (No. RS-2023-00210018) and the Korea Institute of Energy Technology Evaluation and Planning (KETEP) and the Ministry of Trade, Industry and Energy (MOTIE) of the Republic of Korea (No. RS-2023-00236325)}
}

\begin{document}

\IEEEoverridecommandlockouts
\maketitle
\IEEEpubidadjcol

\begin{abstract}
    The increase in renewable energy sources (RES) has reduced power system inertia, making frequency stabilization more challenging and highlighting the need for fast frequency response (FFR) resources. While building energy management systems (BEMS) equipped with distributed energy resources (DERs) can provide FFR, individual BEMS alone cannot fully meet demand. To address this, we propose a community energy management system (CEMS) operational model that minimizes energy costs and generates additional revenue, which is provided FFR through coordinated DERs and building loads under photovoltaic (PV) generation uncertainty. The model incorporates a hierarchical control framework with three levels: Level 1 allocates maximum FFR capacity, Level 2 employs scenario-based stochastic model predictive control (SMPC) to adjust DER operations and ensure FFR provision despite PV uncertainties, and Level 3 performs rapid load adjustments in response to frequency fluctuations detected by a frequency meter. Simulation results on a campus building cluster demonstrate the effectiveness of the proposed model, achieving a 10\% reduction in energy costs and a 24\% increase in FFR capacity, all while maintaining occupant comfort and enhancing frequency stabilization.
\end{abstract}

\section{Introduction}\label{Sec:Intro}
Efforts toward deep de-carbonization of power systems have led to the rapid growth of renewable energy sources (RES). However, the intermittent nature of RES introduces new challenges in maintaining power system security \cite{ipcc2021re}. Because RES reduces system inertia, \jjs{compromising frequency stability.} Consequently, securing fast frequency response (FFR) resources has become essential \cite{nerc2020ffr}. FFR resources, which respond to frequency deviations more swiftly than traditional ancillary services, are key elements in stabilizing power systems with high RES penetration. 

Several countries have presented requirements for FFR, responses within a few seconds and a minimum capacity. AESO operates FFR resources with a 0.2 second response time, ERCOT requirements range from 0.25 to 0.5 seconds, AEMO deploys FFR services within 2 seconds, EirGrid provides reserves up to 75 MW for deviations between 49.85 and 49.3 Hz \cite{FFR_1, FFR_2, FFR_5, FFR_7}. Specifically, PJM provides RegD signal to enable market participants to engage in competitive bidding \cite{FFR_3}. To meet these criteria, a method utilizing Building energy management systems (BEMS) to provide fast demand response at the load level is being proposed \cite{BEMS_2}.

BEMS optimize energy usage within \js{DERs} such as heating, ventilation and air conditioning (HVAC) systems, photovoltaic (PV), electric vehicles (EV), and energy storage systems (ESS); managing demand and supply in real time to improve efficiency and provide ancillary services like demand response \cite{BEMS_1}. However, minimizing operational costs or providing FFR through coordination among DERs is challenging with BEMS alone \cite{commu}. Therefore, we propose a community energy management system (CEMS) that forms a community by integrating BEMS from buildings with different loads and DERs.

HVAC and ESS in CEMS are key DERs capable of providing FFR \cite{HVAC_3, ESS_7}. The study in~\cite{HVAC_3} builds upon a first order equivalent thermal parameter (ETP) model. The ESS model from \cite{ESS_7}, which uses the state of charge (SoC) for building load and FFR provision, enhance to include real-time SoC updates. By integrating both ESS and HVAC into the CEMS model and incorporating equations to calculate an occupants comfort based on indoor temperature \cite{discom}, our study aims to manage building load while reliably providing FFR.

Operating CEMS involves uncertainties such as weather and PV generation \cite{BEMS_33,BEMS_30}. In \cite{HVAC_1}, a hierarchical control method addresses frequency deviation uncertainty for FFR but has limitations regarding PV uncertainty. Building on this approach, our study incorporates scenario-based stochastic model predictive control (SMPC) to better handle uncertainties. SMPC accounts for probabilistic characteristics, generating various scenarios and optimizing the operation of DERs to address PV generation uncertainties and other variables. This approach considers DER operation strategies and PV uncertainty, providing a CEMS operation model that maximizes FFR capacity and effectively responds to real-time frequency deviations for building operators.

This paper addresses the aforementioned limitations and makes the following contributions:
\begin{itemize}
    \item An energy-sharing strategy is proposed among buildings within CEMS, highlighting the contrast with traditional BEMS to enable coordinated demand and supply management, thereby reducing operational costs.
    \item We use a three-level hierarchical control framework, which incorporates frequency metering, FFR provision, and occupant discomfort management, expanding the role of DERs within buildings to include both load management and FFR for frequency stabilization. 
    \item SMPC is introduced to address the uncertainty in PV generation, ensuring reliable energy management and enhancing power system security.
\end{itemize}

\section{Model Formulation}\label{Sec:Model}
\vspace{-1mm}
\subsection{Community Management Structure}
\begin{figure}[t] 
    \centering  
    \includegraphics[width=0.5\textwidth]{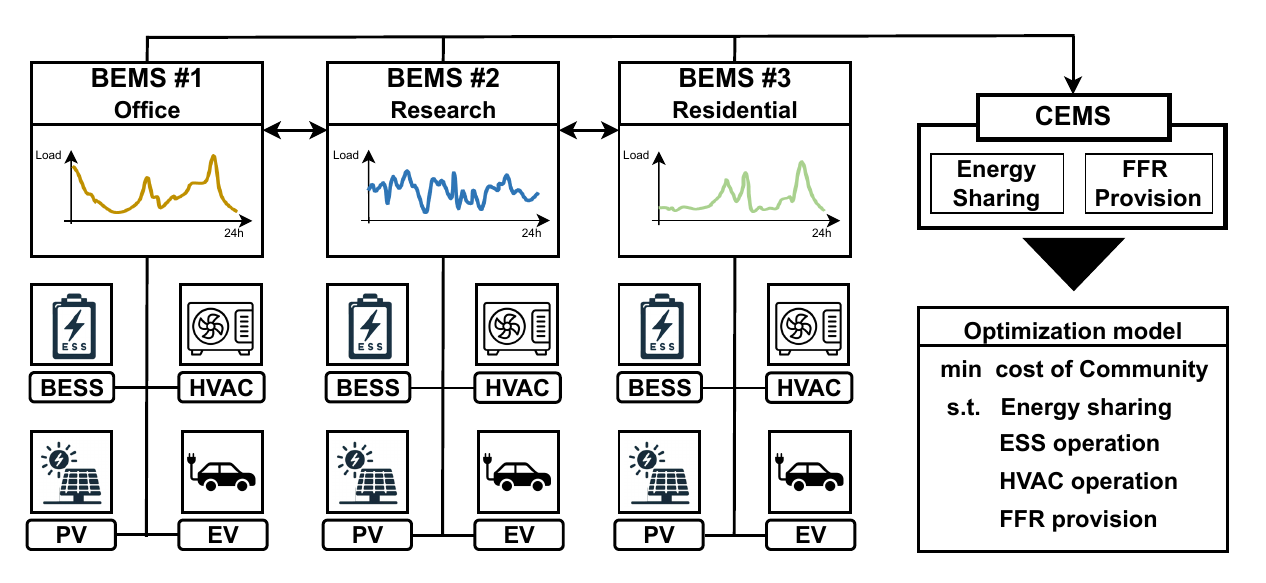}  
    \vspace{-7mm}
    \caption{\small Provision of FFR through community structure and energy sharing between buildings}  
    \vspace{-6mm}
    \label{fig:CEMS_opt}  
\end{figure}

\vspace{-1mm}
CEMS consists of office, research, and residential buildings, each with unique load characteristics and DERs, including ESS, HVAC, PV, and EV, as shown in Fig.~\ref{fig:CEMS_opt}. These DERs are managed using optimization techniques, incorporating specific variables for handling building loads and providing up and down reserves for FFR. The PV system operates using Maximum Power Point Tracking (MPPT) control, and only the charging load is considered for EV. The power balance facilitates energy sharing and optimal DER utilization among buildings, enhancing resource efficiency within the community. Consequently, all resources within the CEMS are coordinated for all time steps ($t \in \mathcal{T}$), buildings ($b \in \mathcal{B}$), zones ($i \in \mathcal{I}$), and scenarios ($s \in \mathcal{S}$).

\vspace{-1mm}
\subsection{PV Generation Modeling}
\vspace{-1mm}
The photovoltaic array output power at the maximum power point can be determined as a linear function of the solar irradiance level. The MPPT mechanism ensures the array operates at its optimal point to maximize power generation as follow \cite{mppt}:
\begin{align}
    p^{\mathrm{PV}}_{t,b,s} = P^{\mathrm{PV,max}}\frac{G_{t,s}}{1000}\eta^{\mathrm{PV}}, \quad \forall t \in \mathcal{T}, b \in \mathcal{B}, s \in \mathcal{S} \label{eq:mppt}
\end{align} 
where \( p_{t,b}^{\mathrm{PV}} \) represents the output power of the PV in \eqref{eq:mppt}, while \( P^{\mathrm{PV, max}} \) is the maximum power output of the PV under standard conditions. The term $G_{t,s}$ denotes the solar irradiance in W/m², with \( 1000 \) being the reference irradiance level \js{and \( \eta^{\mathrm{PV}} \) represents the efficiency of the PV systems.}
\vspace{-1mm}

\subsection{Energy Sharing and Power Balance}
\vspace{-1mm}
The power balance equation ensures that power generated or imported within CEMS matches power consumed or exported, enhancing the reliability of CEMS operations, as follows:
\begin{align}
    &p^{\mathrm{pv}}_{t,b,s} + p^{\mathrm{dis}}_{t,b} + p^{\mathrm{im}}_{t,b} = p^{\mathrm{ch}}_{t,b} + p^{\mathrm{h}}_{t,b,i} + p^{\mathrm{ex}}_{t,b} + D^{\mathrm{ev}}_{t,b} + D^{\mathrm{l}}_{t,b}, \notag \\
    &\quad \forall t \in \mathcal{T}, b \in \mathcal{B}, s \in \mathcal{S} 
    \label{eq:es}
\end{align}
where~\eqref{eq:es} represents the power balance within the CEMS. The left-hand side consists of PV generation \( p^{\mathrm{pv}}_{t,b,s} \), ESS discharge \( p^{\mathrm{dis}}_{t,b} \), and imported power \( p^{\mathrm{im}}_{t,b} \). The right-hand side includes ESS charging \( p^{\mathrm{ch}}_{t,b} \), HVAC power \( p^{\mathrm{h}}_{t,b,i} \), exported power \( p^{\mathrm{ex}}_{t,b} \), EV demand \( D^{\mathrm{ev}}_{t,b} \), and building load demand \( D^{\mathrm{l}}_{t,b} \). This equation ensures balanced power flow across the CEMS, enabling efficient use of energy resources within the community.

\vspace{-1mm}
\subsection{Residential Thermal Modeling}
\vspace{-1mm}
HVAC systems are modeled using the ETP model \cite{HVAC_3}, with additional discomfort constraints incorporated based on the discomfort modeling approach proposed in \cite{discom}, as follows:

\vspace{-5mm}
\begin{align}
    &\underline{T}^{\mathrm{in}} \leq t^{\mathrm{in}}_{t,b,i} \leq \overline{T}^{\mathrm{in}}, \quad \forall t \in \mathcal{T}, b \in \mathcal{B}, i \in \mathcal{I} \label{eq:hvac_14} \\
    &t^{\mathrm{in}}_{t+1,b,i} = 
    \left( 1 - \frac{1}{C_{b,i} R_{b,i}} \right) t^{\mathrm{in}}_{t,b,i} \notag \\
    &+ \frac{1}{R_{b,i}} \left( T^{\mathrm{out}}_t - \eta^{\mathrm{h}}_{b,i} p^{\mathrm{h}}_{t,b,i} \right), 
    \; \forall t \in \mathcal{T}, b \in \mathcal{B}, i \in \mathcal{I} \label{eq:hvac_16} \\
    &\left( 1 - \frac{1}{C_{b,i} R_{b,i}} \right) t^{\mathrm{in}}_{t,b,i} 
    + \frac{1}{R_{b,i}} \left( T^{\mathrm{out}}_t - \eta^{\mathrm{h}}_{b,i} r^{\mathrm{h,dn}}_{t,b,i} \right) 
    \leq \overline{T}^{\mathrm{in}}, \notag \\
     &\quad \forall t \in \mathcal{T}, b \in \mathcal{B}, i \in \mathcal{I} \label{eq:hvac_17} \\
    &\left( 1 - \frac{1}{C_{b,i} R_{b,i}} \right) t^{\mathrm{in}}_{t,b,i} 
    + \frac{1}{R_{b,i}} \left( T^{\mathrm{out}}_t - \eta^{\mathrm{h}}_{b,i} r^{\mathrm{h,up}}_{t,b,i} \right) 
    \geq \underline{T}^{\mathrm{in}}, \notag \\
     &\quad \forall t \in \mathcal{T}, b \in \mathcal{B}, i \in \mathcal{I} \label{eq:hvac_18} \\
    &\sigma^{\mathrm{}}_{t,b,i} \geq 0.01087 \, {t^{\mathrm{in}}_{t,b,i}}^2 - 0.5541 \, t^{\mathrm{in}}_{t,b,i} + 6.8587, \notag \\
    &\quad \forall t \in \mathcal{T}, b \in \mathcal{B}, i \in \mathcal{I} \label{eq:hvac_19}     
\end{align}
where \( \underline{T}^{\mathrm{in}} \) and \( \overline{T}^{\mathrm{in}} \) represent the lower and upper bounds of the acceptable indoor temperature range. The indoor temperature \( t^{\mathrm{in}}_{t,b,i} \) is maintained within these bounds by~\eqref{eq:hvac_14}. The thermal dynamics of the building zone are governed by~\eqref{eq:hvac_16}, which depends on the current indoor temperature \( t^{\mathrm{in}}_{t,b,i} \), outdoor temperature \( T^{\mathrm{out}}_t \), and HVAC system power \( p^{\mathrm{h}}_{t,b,i} \). This evolution is influenced by the thermal resistance \( R_{b,i} \), heat capacity \( C_{b,i} \), and HVAC system efficiency \( \eta^{\mathrm{h}}_{b,i} \). To secure up and down FFR capacity, \( r^{\mathrm{h,up}}_{t,b,i} \) and \( r^{\mathrm{h,dn}}_{t,b,i} \), \eqref{eq:hvac_17} and \eqref{eq:hvac_18} impose additional constraints that ensure \( t^{\mathrm{in}}_{t,b,i} \) respects the comfort range. Occupant comfort \( \sigma^{\mathrm{}}_{t,b,i} \) is modeled as a quadratic function of \( t^{\mathrm{in}}_{t,b,i} \) in~\eqref{eq:hvac_19}, penalizing deviations from optimal comfort levels. These equations collectively enable HVAC systems to manage residential environments, balancing occupant comfort, energy efficiency, and FFR capacity.

\vspace{-1mm}
\subsection{ESS Operation Modeling}
\vspace{-1mm}
ESS manages the load of CEMS and provides FFR through charging and discharging operations. The equations that calculate the load of CEMS, as follows \cite{ESS_7}:
\vspace{-1mm}
\begin{align}
    &e_{t+1,b} = e_{t,b} + p_{t,b}^{\mathrm{ch}} \eta_b^{\mathrm{ch}} - p_{t,b}^{\mathrm{dis}} / \eta_b^{\mathrm{dis}}, \quad \forall t \in \mathcal{T}, b \in \mathcal{B}  \label{eq:ess_energy} \\
    &0.2 E^{\mathrm{}}_b \leq e_{t,b} \leq 0.8 E^{\mathrm{}}_b, \quad \forall t \in \mathcal{T}, b \in \mathcal{B} \label{eq:soc_limits} \\
    &e_{t_0,b} = e_{t_{24},b} = 0.5 E^{\mathrm{}}_b, \quad \forall t \in \mathcal{T}, b \in \mathcal{B} \label{eq:soc_initial_final} \\
    &0 \leq p_{t,b}^{\mathrm{ch}} \leq \overline{P}^{\mathrm{ch}}_b \cdot z_{t,b}^{\mathrm{ch}}, \quad \forall t \in \mathcal{T}, b \in \mathcal{B} \label{eq:charging_limit} \\
    &0 \leq p_{t,b}^{\mathrm{dis}} \leq \overline{P}^{\mathrm{dis}}_b \cdot z_{t,b}^{\mathrm{dis}}, \quad \forall t \in \mathcal{T}, b \in \mathcal{B} \label{eq:discharging_limit} \\
    &z_{t,b}^{\mathrm{ch}} + z_{t,b}^{\mathrm{dis}} \leq 1, \quad \forall t \in \mathcal{T}, b \in \mathcal{B} \label{eq:charging_discharging}
\end{align}
regulate capacity limits and charging/discharging rates. The relationship between the SoC at the current and previous time intervals is governed by~\eqref{eq:ess_energy}. Here, \( e_{t,b} \) denotes the SoC, \( p_{t,b}^{\mathrm{ch}} \) and \( p_{t,b}^{\mathrm{dis}} \) represent the charging and discharging power variables, and \( \eta^{\mathrm{ch}}_{b} \) and \( \eta^{\mathrm{dis}}_{b} \) are their respective efficiencies.  The minimum and maximum SoC limits, defined in~\eqref{eq:soc_limits}, restrict operation to 20\% and 80\% of the maximum capacity. This range prevents excessive degradation by avoiding overcharging or deep discharging, ensuring the longevity and reliability of the ESS. The initial and final SoC are maintained at 50\% in each optimization, as specified in~\eqref{eq:soc_initial_final}. Additionally, the charging and discharging power are constrained by~\eqref{eq:charging_limit} and~\eqref{eq:discharging_limit}, ensuring they remain below the maximum output power \( \overline{P}^{\mathrm{ch}}_b \) and \( \overline{P}^{\mathrm{dis}}_b \), accounting for system efficiency.

The equations for deriving up and down reserves for FFR, considering the SoC, as follows \cite{ESS_7}:
\vspace{-1mm}
\begin{align}
    &e_{t+1,b} \geq e_{t,b} + \eta_b^{\mathrm{ch}} \cdot r_{t,b}^{\mathrm{ch,dn}} - r_{t,b}^{\text{dis,up}} / \eta_b^{\mathrm{dis}}, \; \forall t \in \mathcal{T}, b \in \mathcal{B} \label{eq:ess_7} \\
    &e_{t+1,b} \leq e_{t,b} - \eta_b^{\mathrm{ch}} \cdot r_{t,b}^{\mathrm{ch,up}} + r_{t,b}^{\mathrm{dis,dn}} / \eta_b^{\mathrm{dis}}, \; \forall t \in \mathcal{T}, b \in \mathcal{B} \label{eq:ess_8} \\
    &p_{t,b}^{\mathrm{dis}} + r_{t,b}^{\mathrm{dis,up}} \leq \overline{P}_{b}^{\mathrm{dis}}, \quad \forall t \in \mathcal{T}, b \in \mathcal{B} \label{eq:ess_9} \\
    &p_{t,b}^{\mathrm{dis}} - r_{t,b}^{\mathrm{dis,dn}} \geq \underline{P}_{b}^{\mathrm{dis}}, \quad \forall t \in \mathcal{T}, b \in \mathcal{B} \label{eq:ess_10} \\
    &p_{t,b}^{\mathrm{ch}} - r_{t,b}^{\mathrm{ch,up}} \geq \underline{P}_{b}^{\mathrm{ch}}, \quad \forall t \in \mathcal{T}, b \in \mathcal{B} \label{eq:ess_11} \\
    &p_{t,b}^{\mathrm{ch}} + r_{t,b}^{\mathrm{ch,dn}} \leq \overline{P}_{b}^{\mathrm{ch}}, \quad \forall t \in \mathcal{T}, b \in \mathcal{B} \label{eq:ess_12} \\
    &E_b^{\mathrm{min}} + r_{t,b}^{\mathrm{e,dn}} \leq e_{t,b} \leq E_b^{\mathrm{max}} - r_{t,b}^{\mathrm{e,up}}, \quad \forall t \in \mathcal{T}, b \in \mathcal{B} \label{eq:ess_13}
\end{align}
where ensure reserves are provided within allowable bounds, are governed by~\eqref{eq:ess_7} and \eqref{eq:ess_8}. Here, \( r_{t,b}^{\mathrm{ch,dn}} \) and \( r_{t,b}^{\mathrm{ch,up}} \) represent the charging reserves for down-regulation and up-regulation, respectively, while \( r_{t,b}^{\mathrm{dis,up}} \) and \( r_{t,b}^{\mathrm{dis,dn}} \) denote the discharge reserves. The discharge power, constrained to ensure reliable operation, is defined in~\eqref{eq:ess_9} and \eqref{eq:ess_10}. Specifically, \( p_{t,b}^{\mathrm{dis}} \) is bounded by \( \overline{P}_{b}^{\mathrm{dis}} \) (maximum discharge power) and \( \underline{P}_{b}^{\mathrm{dis}} \) (minimum discharge power). Similarly, the charging power limits are governed by~\eqref{eq:ess_11} and \eqref{eq:ess_12}, where \( p_{t,b}^{\mathrm{ch}} \) is bounded by \( \overline{P}_{b}^{\mathrm{ch}} \) (maximum charging power) and \( \underline{P}_{b}^{\mathrm{ch}} \) (minimum charging power). These constraints enable operational flexibility while supporting FFR. To maintain energy availability for FFR, the energy state \( e_{t,b} \) is kept within the limits defined by~\eqref{eq:ess_13}. Here, \( E_b^{\mathrm{min}} \) and \( E_b^{\mathrm{max}} \) denote the minimum and maximum energy capacities, while \( r_{t,b}^{\mathrm{e,dn}} \) and \( r_{t,b}^{\mathrm{e,up}} \) ensure sufficient reserves for down-regulation and up-regulation, respectively.

\vspace{-1mm}
\section{Hierarchy Framework for CEMS}\label{Sec:Solution}

\subsection{CEMS Control Method}
\vspace{-1mm}
\begin{figure}[t] 
    \centering  
    \includegraphics[width=0.48\textwidth]{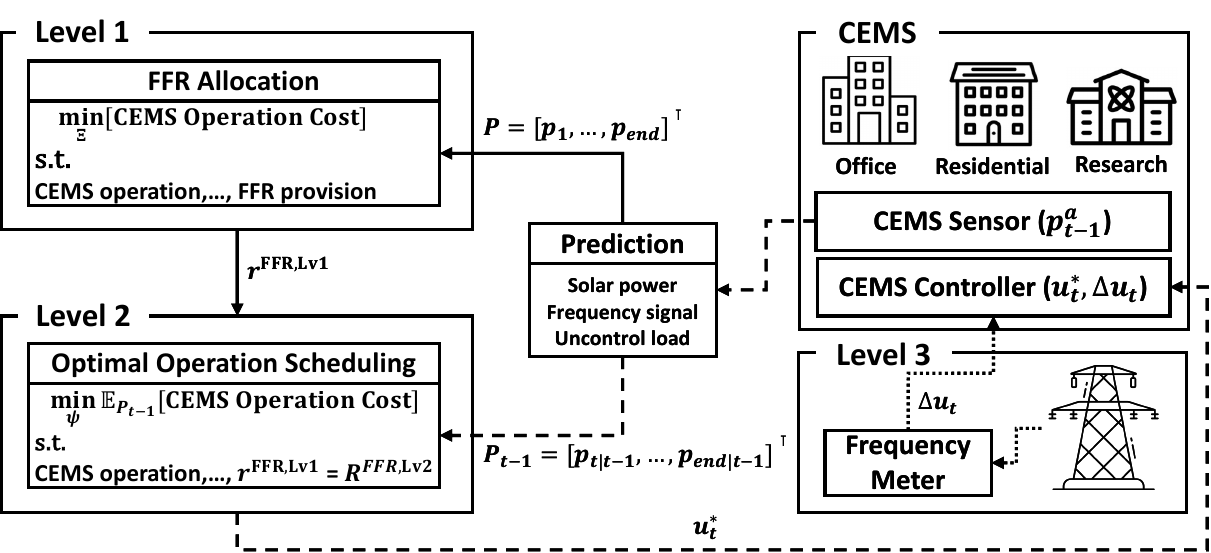}  \vspace{-1mm}
    \caption{\small \jjs{Hierarchical control with SMPC in CEMS: the solid line represents day-ahead Level 1, the dashed line indicates real-time Level 2 with SMPC, and the dotted line shows Level 3 for FFR.}}  
    \vspace{-6.5mm}
    \label{fig:CEMS_cont}  
\end{figure}
Figure~\ref{fig:CEMS_cont} shows the hierarchical control process for CEMS FFR operation. Data from sensors, including PV generation, frequency signals, and uncontrolled loads, informs day-ahead optimization at Level 1 to set FFR capacity. Level 2 applies SMPC to manage real-time uncertainties and adjust DERs to meet the Level 1 target. Level 3 uses frequency meters to address actual deviations, dynamically adjusting DERs for precise FFR provision.

\vspace{-1mm}
\subsection{Level 1: FFR Allocation}
\vspace{-1mm}
Using the most probable PV generation forecast scenario from day ahead, the optimization problem for minimizing CEMS operating costs while determining the FFR capacity is as follows:
\begin{align}
    &\min_{\Xi} \sum_{t \in \mathcal{T}} \sum_{b \in \mathcal{B}} 
    \bigg[ \lambda^{\mathrm{im}}_{t,b} p^{\mathrm{im}}_{t,b} 
    - \lambda^{\mathrm{ex}}_{t,b} p^{\mathrm{ex}}_{t,b} \notag \\
    &\quad - \sum_{i \in \mathcal{I}} \Big\{\lambda^{\mathrm{comf}}_{t,b,i} \sigma_{t,b,i}
    - \lambda^{\mathrm{ffr}} \big( r^{\mathrm{up}}_{t,b,i} + r^{\mathrm{dn}}_{t,b,i} \big) 
    \Big\}\bigg], \label{eq:lv1_obj} \\
    &\text{s.t.} \quad \eqref{eq:mppt} - \eqref{eq:ess_13}, \label{eq:lv1_st} \\
    &r^{\mathrm{up}}_{t,b,i} = r^{\mathrm{e,up}}_{t,b} + r^{\mathrm{h,up}}_{t,b,i}, 
    \quad \forall t \in \mathcal{T}, b \in \mathcal{B}, i \in \mathcal{I} \label{eq:lv1_up} \\
    &r^{\mathrm{dn}}_{t,b,i} = r^{\mathrm{e,dn}}_{t,b} + r^{\mathrm{h,dn}}_{t,b,i}, 
    \quad \forall t \in \mathcal{T}, b \in \mathcal{B}, i \in \mathcal{I} \label{eq:lv1_dn}
\end{align}
where $\Xi \coloneqq \big[p^{\mathrm{im}}_{t,b}, p^{\mathrm{ex}}_{t,b}, p^{\mathrm{ch}}_{t,b}, p^{\mathrm{dis}}_{t,b}, e_{t,b}, p^{\mathrm{h}}_{t,b}, r^{\mathrm{up}}_{t,b,i}, r^{\mathrm{dn}}_{t,b,i}, z_{t,b} \in \{0,1\}, \sigma_{t,b,i} \in \mathbb{R}\big]$ represents the set of decision variables for Level~1 optimization. The objective function in~\eqref{eq:lv1_obj} aims to minimize total system costs. Here, \( p^{\mathrm{im}}_{t,b} \) and \( p^{\mathrm{ex}}_{t,b} \) denote the import and export power, respectively, with their associated costs represented by \( \lambda^{\mathrm{im}}_{t,b} \), \( \lambda^{\mathrm{ex}}_{t,b} \), and \( \lambda^{\mathrm{disc}}_{t,b,i} \). The term \( \sigma_{t,b,i} \) accounts for occupant comfort, while \( \lambda^{\mathrm{ffr}} \) denotes the revenue generated from FFR. The variables \( r^{\mathrm{up}}_{t,b,i} \) and \( r^{\mathrm{dn}}_{t,b,i} \) represent the up and down reserves for FFR. The constraints in \eqref{eq:lv1_st} ensure that the optimization problem satisfies conditions \eqref{eq:mppt}-\eqref{eq:ess_13}. The up and down reserves, provided by the ESS and HVAC systems, are defined in \eqref{eq:lv1_up} and \eqref{eq:lv1_dn}. These reserves ensure the efficient allocation of FFR across the system.

\vspace{-1mm}
\subsection{Level 2: Optimal Operation Scheduling}
\vspace{-1mm}
By applying SMPC considering the probability of PV generation at each time step, the control variables are adjusted to achieve the target FFR capacity determined in Level 1, as follows:
\begin{align}
    &\min_{\Psi} \sum_{t = k}^{k+N-1} \sum_{b \in \mathcal{B}} \sum_{s \in \mathcal{S}} 
    \Bigg[ \omega_s \Bigg\{ \lambda^{\mathrm{im}}_{t,b} p^{\mathrm{im}}_{t,b}
    - \lambda^{\mathrm{ex}}_{t,b} p^{\mathrm{ex}}_{t,b} \notag \\
    &\quad - \sum_{i \in \mathcal{I}} \lambda^{\mathrm{comf}}_{t,b,i} \sigma_{t,b,i} \Bigg\}
    - \sum_{i \in \mathcal{I}} \lambda^{\mathrm{ffr}} \big( R^{\mathrm{up}}_{t,b,i} + R^{\mathrm{dn}}_{t,b,i} \big) \Bigg], \label{eq:lv2_obj} \\
    &\text{s.t.} \quad \eqref{eq:mppt}-\eqref{eq:ess_13}, \label{eq:lv2_st} \\
    &R^{\mathrm{e,up}}_{t,b} = r^{\mathrm{e,up}}_{t,b}, 
    \quad \forall t \in [k, k+N-1], b \in \mathcal{B} \label{eq:lv2_eup} \\ 
    &R^{\mathrm{e,dn}}_{t,b} = r^{\mathrm{e,dn}}_{t,b}, 
    \quad \forall t \in [k, k+N-1],  b \in \mathcal{B} \label{eq:lv2_edn} \\
    &R^{\mathrm{h,up}}_{t,b,i} = r^{\mathrm{h,up}}_{t,b,i}, 
    \quad \forall t \in [k, k+N-1], b \in \mathcal{B}, i \in \mathcal{I} \label{eq:lv2_hup} \\
    &R^{\mathrm{h,dn}}_{t,b,i} = r^{\mathrm{h,dn}}_{t,b,i}, 
    \quad \forall t \in [k, k+N-1], b \in \mathcal{B}, i \in \mathcal{I} \label{eq:lv2_hdn}
\end{align}
where $\Psi \coloneqq \big[p^{\mathrm{im}}_{t,b}, p^{\mathrm{ex}}_{t,b}, p^{\mathrm{ch}}_{t,b}, p^{\mathrm{dis}}_{t,b}, e_{t,b}, p^{\mathrm{h}}_{t,b}, z_{t,b} \in \{0,1\}, \sigma_{t,b,i} \in \mathbb{R}\big]$ represents the set of decision variables for Level~2. The objective function in~\eqref{eq:lv2_obj} minimizes CEMS operation costs by incorporating the probabilities of scenarios in the set \( \mathcal{S} \). Scenarios are generated to account for uncertainty in PV generation, as defined by~\eqref{eq:mppt}. The constraints in~\eqref{eq:lv2_st} ensure that the FFR capacity determined in Level~1 is achieved for all scenarios. The up and down reserves provided by DERs are defined in~\eqref{eq:lv2_eup}-\eqref{eq:lv2_hdn}, where the reserves from ESS are denoted as \( R^{\mathrm{e,up}}_{t,b} \) for up FFR and \( R^{\mathrm{e,dn}}_{t,b} \) for down FFR, and the reserves from HVAC systems are denoted as \( R^{\mathrm{h,up}}_{t,b,i} \) for up FFR and \( R^{\mathrm{h,dn}}_{t,b,i} \) for down FFR. The values \( r^{\mathrm{up}}_{t,b,i} \) and \( r^{\mathrm{dn}}_{t,b,i} \) determined in Level~1 are fixed as \( R^{\mathrm{e,up}}_{t,b} \), \( R^{\mathrm{e,dn}}_{t,b} \), \( R^{\mathrm{h,up}}_{t,b,i} \), and \( R^{\mathrm{h,dn}}_{t,b,i} \) in Level~2 to ensure consistency with the optimized FFR capacity.

\vspace{-1mm}
\subsection{Level 3: Frequency Control}
\vspace{-1mm}
The goal of Level 3 is to provide FFR in response to frequency fluctuations, defined as:
\begin{align}
    &u_{t,b}^{\mathrm{Lv3}} = u_{t,b}^{\mathrm{Lv2}} + \Delta u_{t,b} = u_{t,b}^{\mathrm{Lv2}} + w_t \big(r^{\mathrm{up}}_{t,b,i} + r^{\mathrm{dn}}_{t,b,i} \big), \notag \\
    &\quad \text{where} \quad u_{t,b}^{\mathrm{Lv2}} \in \Psi, 
    \quad \forall t \in \mathcal{T}, b \in \mathcal{B}, i \in \mathcal{I} \label{eq:lv3}
\end{align}
where \eqref{eq:lv3} defines the final control input \( u_{e,t}^{\mathrm{Lv3}} \), which is calculated by adding a correction term \( \Delta u_{e,t} \) to the Level 2 control input \( u_{e,t}^{\mathrm{Lv2}} \). This correction, determined by frequency deviation, utilizes up and down reserves \( r^{\mathrm{up}}_{t,b,i} \), \( r^{\mathrm{dn}}_{t,b,i} \) weighted by \( w_t \). Level 3 dynamically adjusts the Level 2 control input (\( u_{e,t}^{\mathrm{Lv2}} \in \Psi \)) to respond in real time.

\vspace{-1mm}
\section{Case Study}\label{Sec:case}
\vspace{-1mm}
\subsection{Basic data}
\vspace{-1mm}
\begin{figure}[t]  
    \centering  
    \begin{subfigure}[b]{0.48\textwidth}  
        \includegraphics[width=\textwidth]{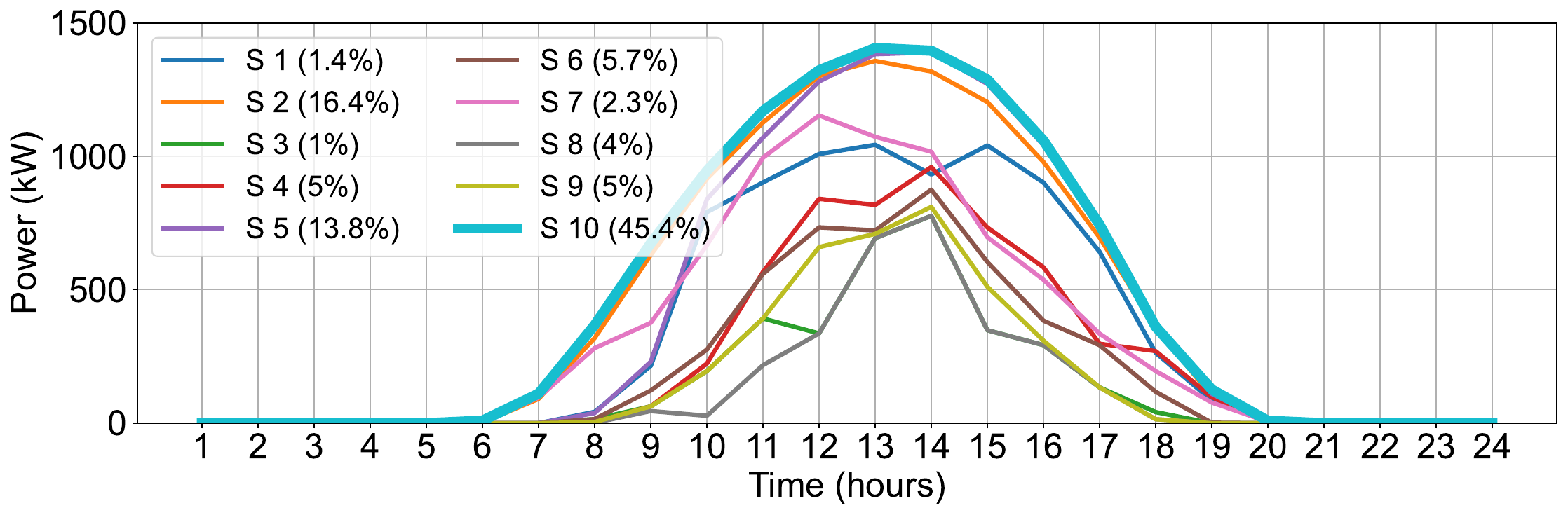}
        \vspace{-6mm}
        \caption{\small PV generation scenarios in Level 1 (S 10)}
        \label{fig:pv_gen}
    \end{subfigure}
    \begin{subfigure}[b]{0.48\textwidth}
        \includegraphics[width=\textwidth]{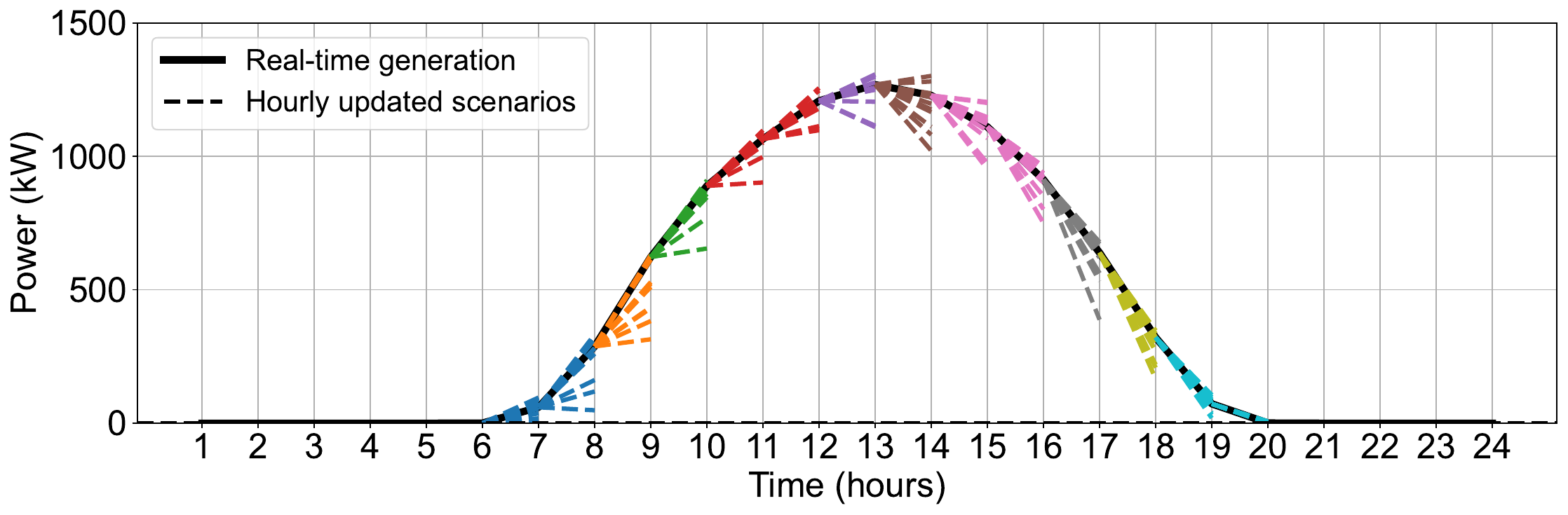}
        \vspace{-6mm}
        \caption{\small PV generation scenarios in Level 2 (Hourly updated)}
        \label{fig:pv_gen_2}
    \end{subfigure}
    \vspace{-2mm}
    \caption{\small PV generation}  
    \vspace{-7mm}
    \label{fig:pv} 
\end{figure}
Using the methodology from \cite{pv_sr}, 10 scenarios were generated in in Fig.~\ref{fig:pv_gen}. Building on this, we developed a scenario update method using a covariance matrix capturing solar power variability in Fig.~\ref{fig:pv_gen_2}. The frequency signal was derived from the RegD test signal \cite{PJM_RegD_2024}. HVAC system parameters, including heat capacity, thermal resistance, and system efficiency, were adjusted according to building characteristics \cite{HVAC_4}. ESS units with a capacity of 1 MW were installed in each building, with charging and discharging efficiencies set at 90\% and 80\% \cite{ESS_7}. A total of 10 MW of PV capacity was installed. Electricity prices followed KEPCO retail rates \cite{kepco_rates}.
\begin{figure}[t] 
    \centering  
    \begin{subfigure}[b]{0.48\textwidth}  
        \includegraphics[width=\textwidth]{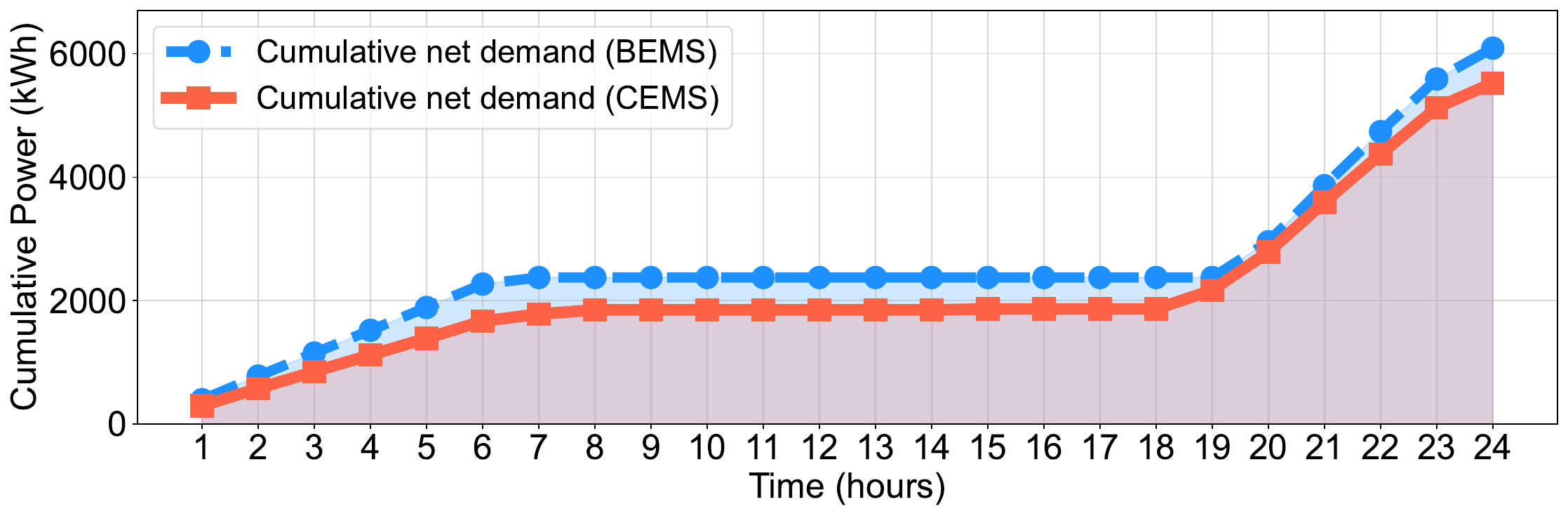}
        \vspace{-6mm}
        \caption{\small Net demand}
        \label{fig:energy_sharing}
    \end{subfigure}
    \begin{subfigure}[b]{0.48\textwidth}
        \includegraphics[width=\textwidth]{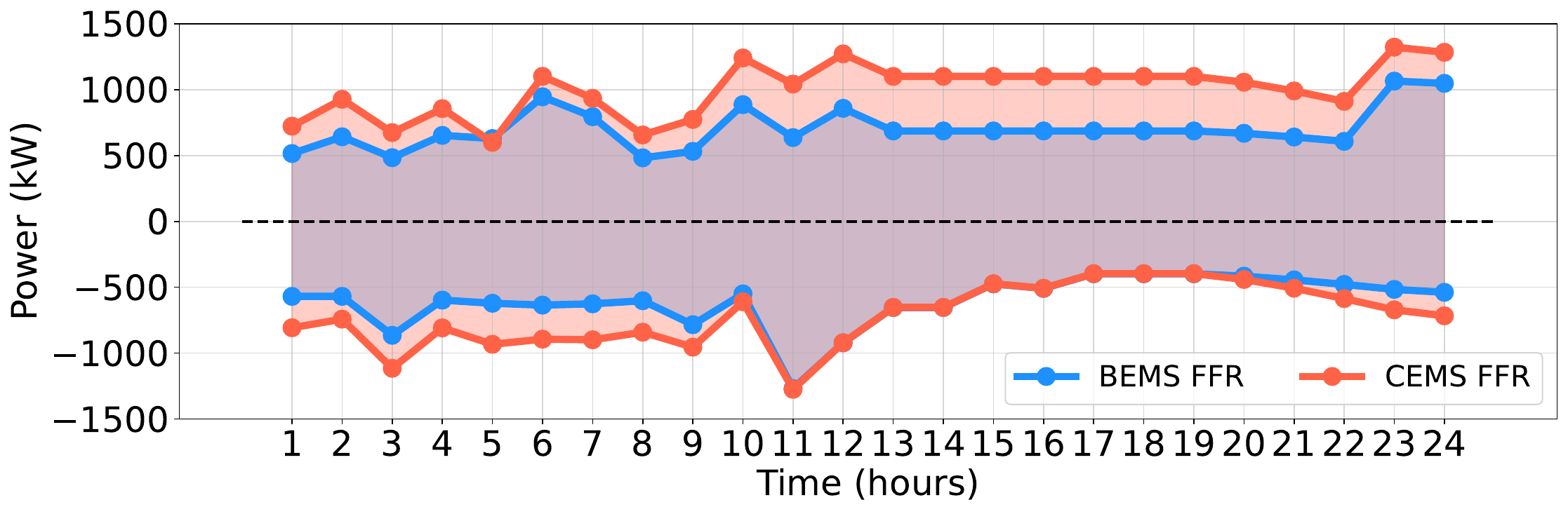}
        \vspace{-6mm}
        \caption{\small FFR capacity}
        \label{fig:FFR_CEMS_BEMS}
    \end{subfigure}
    \vspace{-2mm}
    \caption{\small Community net demand with energy sharing}  
    \vspace{-3mm}
    \label{fig:energy_sharing_ffr} 
\end{figure}
\vspace{-6mm}
\subsection{Energy Sharing and FFR Provision}
\begin{figure}[t] 
    \centering  
    \begin{subfigure}[b]{0.48\textwidth}  
        \includegraphics[width=\textwidth]{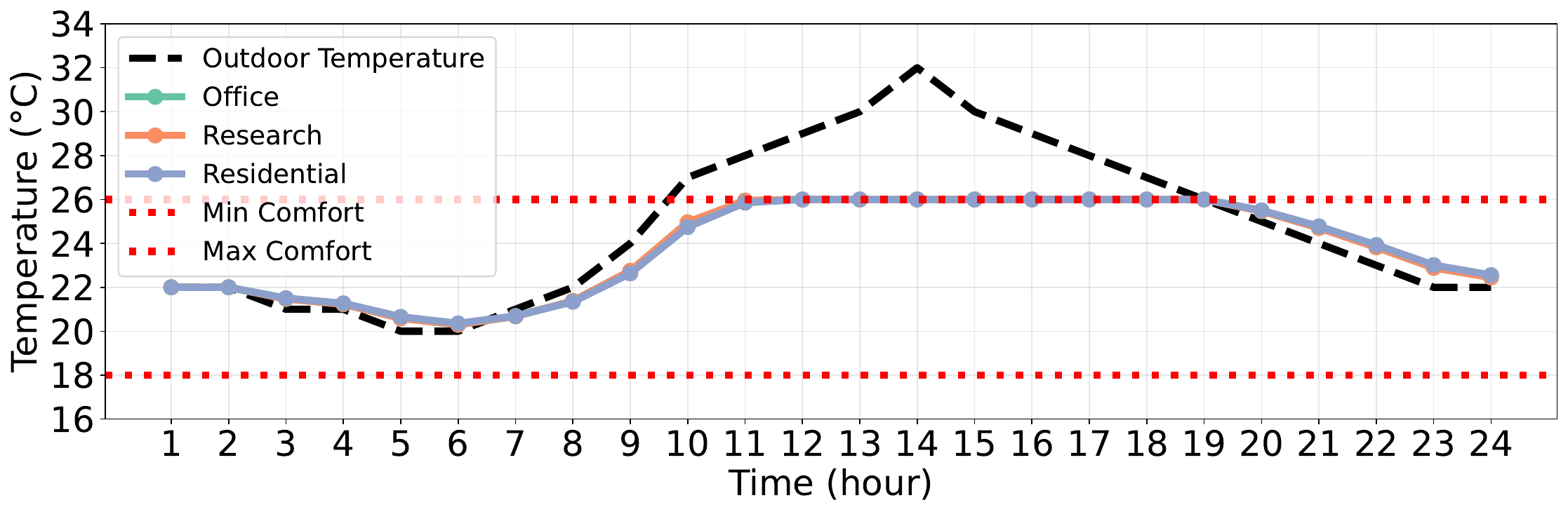}
        \vspace{-6.3mm}
        \caption{\small Indoor and outdoor temperature}
        \label{fig:in_out_temp}
    \end{subfigure}
    \begin{subfigure}[b]{0.48\textwidth}
        \includegraphics[width=\textwidth]{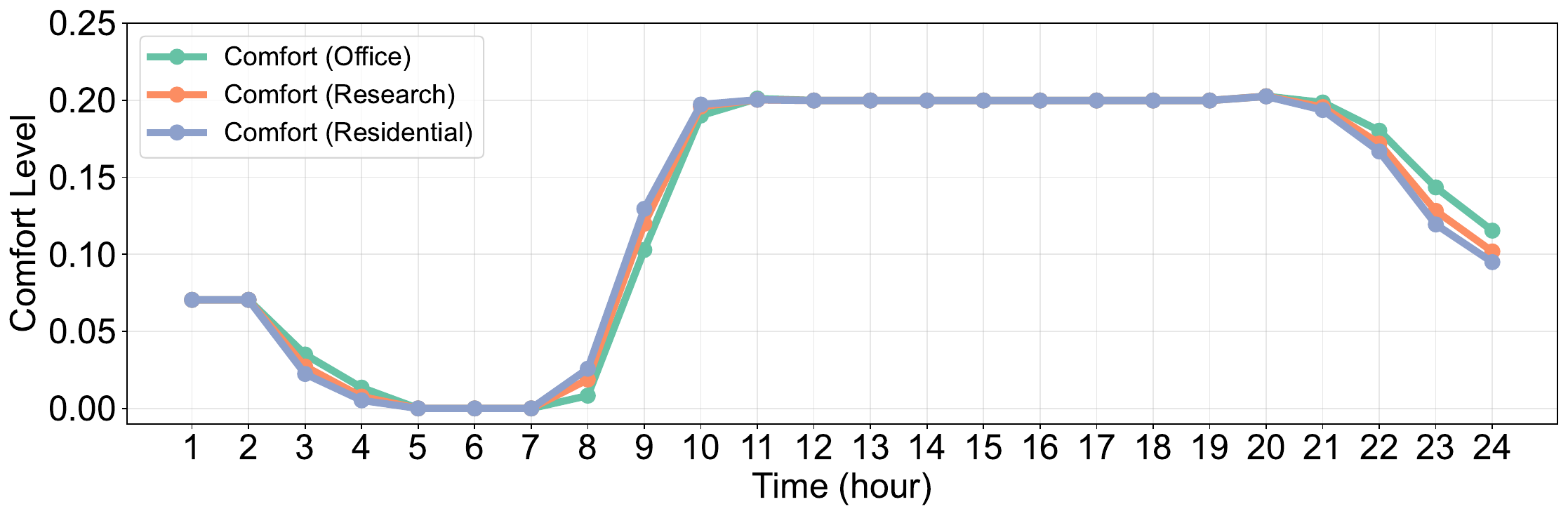}
        \vspace{-6.3mm}
        \caption{\small Occupants comfort level}
        \label{fig:comf}
    \end{subfigure}
    \begin{subfigure}[b]{0.48\textwidth}
        \includegraphics[width=\textwidth]{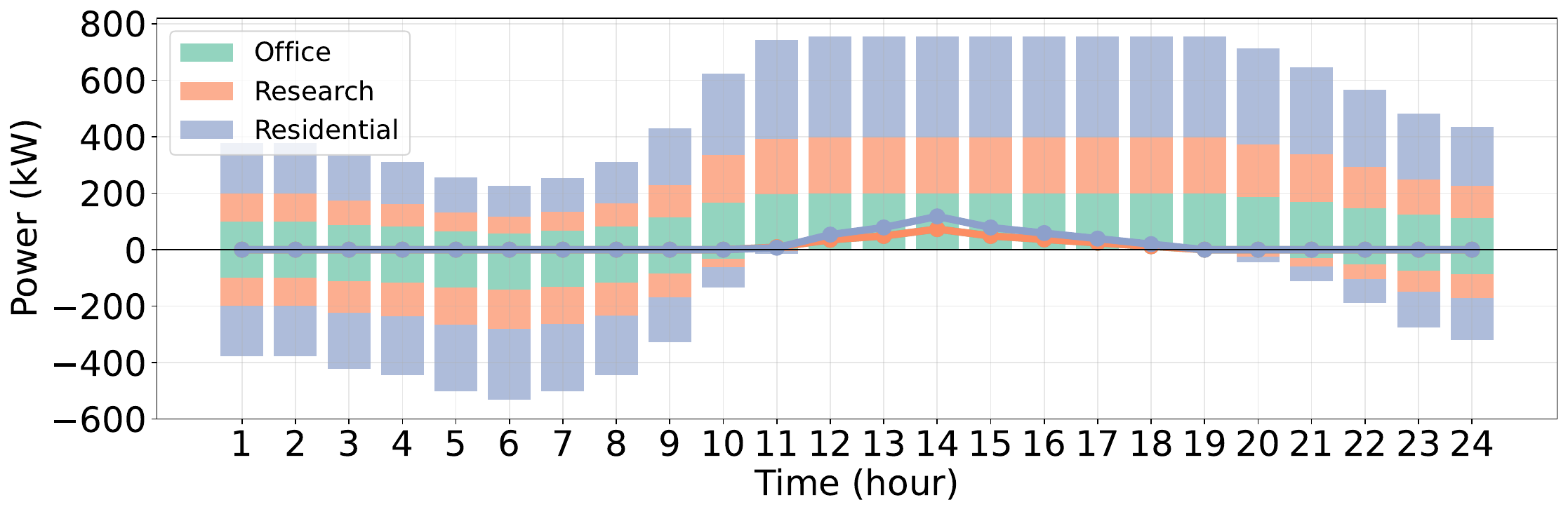}
        \vspace{-6.3mm}
        \caption{\small HVAC power consumption and FFR capacity}
        \label{fig:hvac_cems_op}
    \end{subfigure}
    \vspace{-1.3mm}
    \caption{\small HVAC power and FFR capacity analysis for CEMS}  
    \vspace{-4.6mm}
    \label{fig:HVAC_ffr_all}  
\end{figure}

\vspace{-0.35mm}
In BEMS, each building (\( b=1, 2, 3 \)) individually optimizes its operation based on \eqref{eq:mppt} to \eqref{eq:lv1_dn}, whereas CEMS collectively optimizes these equations for all buildings (\( \forall b \)). Figure~\ref{fig:energy_sharing} shows that cumulative net demand is reduced by 10\%. Figure~\ref{fig:FFR_CEMS_BEMS} highlights that CEMS provides 24\% more FFR capacity than BEMS, demonstrating that energy sharing reduces the load burden on individual buildings and enhances FFR provision through effective coordination among DERs.
\begin{figure}[t] 
    \centering  
    \includegraphics[width=0.48\textwidth]{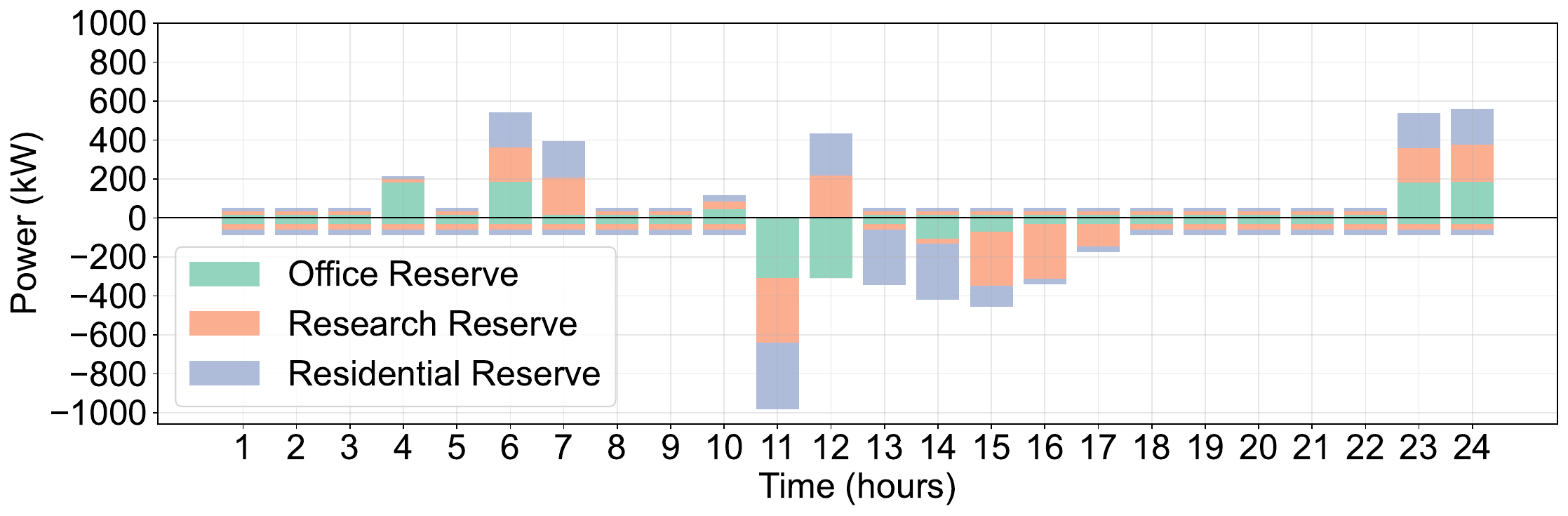}  
    \vspace{-2.5mm}
    \caption{\small ESS operation and reserve analysis for CEMS}  % 그림에 대한 설명
    \label{fig:ess_ffr_all} 
    \vspace{-6mm}
\end{figure}
Figure~\ref{fig:ess_ffr_all} presents the modeling of ESS reserves to provide consistent reserve capacity throughout the 24-hour period based on the state of charge (SoC). Between 1:00 and 7:00, when building loads are low, charging occurs, enabling the ESS to secure upward reserve capacity during this time. During periods of increased photovoltaic generation, the ESS discharges the energy stored earlier to reduce the net demand of buildings. At 12:00, both the research and residential buildings begin recharging while the office ESS continues to discharge, preparing for subsequent discharging needs in later hours. Finally, between 23:00 and 24:00, the ESS maintains an SoC of approximately 50\%, adhering to the constraint defined in \ref{eq:soc_initial_final}. This result highlights the effectiveness of ESS in simultaneously managing building loads and providing FFR.
Figure~\ref{fig:HVAC_ffr_all} shows the HVAC operation and reserve analysis for each building under CEMS. Indoor temperatures, as shown in Fig.~\ref{fig:in_out_temp}, remain within the comfort range of 18°C to 26°C (dashed red lines). Fig.~\ref{fig:hvac_cems_op} illustrates HVAC power consumption, with the research building peaking at 150 kW, the office at 100 kW, and the residential building at 50 kW. These variations stem from differences in thermodynamic properties. The bar graphs indicate that from 11:00 to 19:00, HVAC systems cannot provide down FFR due to high power demand, but approximately 700 kW of FFR is consistently supplied through indoor temperature adjustments. From 9:00 to 19:00, the HVAC operation is set to prioritize occupant satisfaction over FFR revenue. As a result, Fig.~\ref{fig:in_out_temp} and Fig.~\ref{fig:comf} show that, from 10:00 to 19:00, when the outdoor temperature exceeds 26°C, the indoor temperature is maintained at 25.7°C, the level that ensures the highest occupant satisfaction.
\vspace{-1.5mm}
\subsection{SMPC and Hierarchical Control for Uncertainty}
\begin{figure}[t]  
    \centering  
    \includegraphics[width=0.48\textwidth]{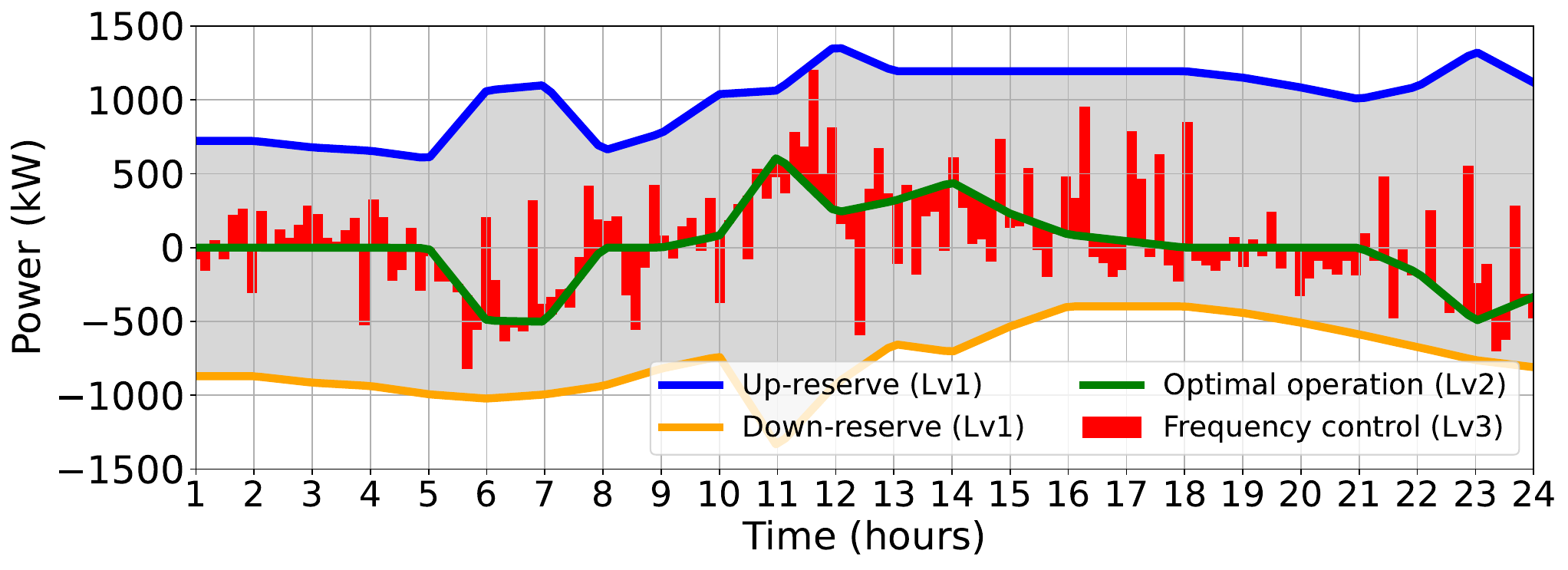}  
    \vspace{-3mm}
    \caption{\small Hierarchical control of FFR for CEMS operations}  
    \vspace{-5mm}
    \label{fig:smpc} 
\end{figure}

\vspace{-1mm}
Figure~\ref{fig:smpc} shows the hierarchical control results within the CEMS, combining hierarchical control with SMPC to ensure FFR capacity under PV generation uncertainty. The control structure includes three levels: Level 1 calculates up and down FFR (blue and orange lines) using day-ahead forecasts, Level 2 (green line) optimizes DER scheduling with SMPC to minimize costs and maintain FFR capacity, and Level 3 (red bars) provides frequency control to respond to deviations. This integration efficiently manages PV uncertainty while ensuring FFR capacity.
\vspace{-2mm}
\section{conclusion}\label{Sec:conclu}
This study proposes a hierarchical framework for CEMS that provides FFR while addressing uncertainties in PV generation. The approach comprises three levels: FFR planning, real-time optimization, and DER adjustment for frequency deviations. The results indicate that the model reduces energy costs by 10\% while enhancing FFR provision by 24\% through the coordinated adjustment of DERs and building loads. The simulation for a cluster of campus buildings balanced energy efficiency, occupant comfort, and system reliability, demonstrating the effectiveness of integrating energy sharing, hierarchical control, and SMPC for reliable and cost-effective CEMS operations. Future research will focus on identifying cost-effective operational strategies for providing FFR that balance the interests of EV users and CEMS operators.

\vspace{-1.5mm}

\bibliographystyle{IEEEtran}
\bibliography{ref.bib}

\end{document}